\documentclass{aa}
\usepackage{graphicx}
\usepackage[round]{natbib}
\begin{document}

\title{Electrostatic Instability in Electron-Positron Pairs 
Injected in an External Electric Field}

\author{Katsuaki Asano \and Fumio Takahara}

\institute{Department of Earth and Space Science,
Osaka University, Toyonaka 560-0043, Japan\\
\email{asano@vega.ess.sci.osaka-u.ac.jp, 
       takahara@vega.ess.sci.osaka-u.ac.jp}}

\titlerunning{Electrostatic Instability in an External Electric Field}
\authorrunning{Asano and Takahara}

\abstract{
Motivated by the particle acceleration problem in pulsars,
we numerically investigate electrostatic instability of 
electron-positron pairs injected in an external electric field.
The electric field is expected to be so strong that
we cannot neglect effects of spatial variation in the 0-th order 
distribution functions on the scale of the plasma oscillation.
We assume that pairs are injected mono-energetically with 4-velocity
$u_0>0$ in a constant external electric field by which electrons 
(positrons) are accelerated (decelerated). By solving linear 
perturbations of the field and distribution functions of pairs, 
we find a new type of electrostatic instability.
The properties of the instability are characterized by $u_0$ and 
the ratio $R$ of the braking time-scale
(determined by the external electric field) to the time-scale
of the plasma oscillation.
The growth rate is as large as a few times the plasma frequency.
We discuss the possibility that
the excited waves prevent positrons from returning to 
the stellar surface.
}

\maketitle


\section{Introducton}

\indent

Waves in homogeneous plasmas are well described by the linear 
perturbations that have the Fourier-harmonic dependence
of the form 
$\exp{[i (\vec{k} \cdot \vec{x}-\omega t)]}$.
The properties of various wave modes have been extensively studied 
for various physical situations.
Plasma instabilities, such as the two-stream instability,
the Weibel instability \citep{wei59}, and many others have been 
recognized as important processes in astrophysics as well as in 
laboratory situations.
For inhomogeneous plasmas, the Fourier-harmonic dependence
is not assured in a strict sense.
If the wavelength is short enough compared with the scale
of the inhomogeneities, we may neglect effects of the spatial gradient
of the distribution functions of particles, and carry
out the Fourier-harmonic expansion.
We call this treatment the local approximation hereafter.

When static electric field exists in a plasma,
it accelerates particles and leads to inhomogeneous velocity 
distributions of particles.
Wave properties in an electric field were studied
in geophysical researches, adopting the local approximation
(e.g. Misra and Singh 1977, Misra at al. 1979, Das and Singh 1982).
In many high-energy astronomical phenomena, electric field 
is expected to be so strong that the local approximation is not 
adequately applied. Such a situation typically appears in the 
magnetosphere of pulsars. 

A spinning magnetized neutron star provides huge electric
potential differences between different parts of its surface 
as a result of unipolar induction \citep{gol69}.
A part of the potential difference will be expended as an 
electric field along the magnetic field somewhere in the magnetosphere.
Although a fully consistent model for the pulsar magnetosphere has yet 
to be constructed, several promising models have been considered.
Among them, the polar cap model \citep{stu71, rud75} assumes that an electric field 
$E_\parallel$ parallel to the magnetic field lines exists just
above the magnetic poles.
The electric field accelerates charged particles up to TeV energies,
and resultant curvature radiation from these particles produces copious 
electron-positron pairs through magnetic pair production.
These pairs may provide gamma-ray emission by curvature 
radiation or synchrotron radiation as well as coherent radio 
emission and a source for the pulsar wind.

The localized potential drop is maintained by a pair of anode
and cathode regions.
In the cathode region the space charge density $\rho$ deviates
from the Goldreich-Julian (GJ) density 
$\rho_{\rm GJ} \simeq -\Omega B_z/2 \pi c$
negatively, where $\Omega=2 \pi/T$ is the angular velocity of the 
star and $B_z$ is the magnetic field strength along the rotation 
axis. On the other hand, $\rho$ deviates positively for the anode.
Outside the accelerator the electric field is screened out.
In the polar cap model, especially for space charge limited flow model
\citep{faw77, sch78, aro79}, 
where electrons can freely escape from the stellar surface, 
i.e., $E_\parallel = 0$ on the stellar surface,
the formation mechanism of a static pair 
of anode and cathode, which can sustain enough potential drop 
for pair production, is a long-standing issue. 
Current flows steadily along the magnetic field line so that 
the charge density is determined by the magnitude of the current 
and field geometry with suitable boundary conditions. 
Good examples for space charge limited flow 
are in \citet{shi97}.
When $\rho_{\rm GJ}<0$ and the electron density ($n \propto B/v \simeq B/c$,
where $B$ is the magnetic field strength) is larger than
the GJ number density ($n_{\rm GJ}=|\rho_{\rm GJ}/e| \propto B_z$,
where $-e$ is the electronic charge)
on the stellar surface, a cathode is provided on
the stellar surface. The cathode accelerates electrons.
When the field lines curve away from the rotation axis, 
$n$ deviates $n_{\rm GJ}$ nagatively so much more for `away' curvature,
which enhances the cathode.
Hence electrons continue to be accelerated,
and potential drop becomes large enough to produce pairs.
The mechanism of the electric field screening, i.e., a way
to provide an anode, has been 
considered to be provided by pair polarization. 
Although most papers take it for granted that copious pair production
can instantly screen the field, recently \citet{shi98, shi02} casted 
doubt on this issue; the electric field screening is not an 
easy task as considered usually.

\citet{shi98, shi02} investigated the screening of electric fields
in the pair production region. 
They found that the thickness of the screening layer is restricted to be
as small as the braking distance $l_{\rm E}=m_{\rm e} c^2/|e E_\parallel|$
for which decelerating particles become non-relativistic,
where $m_{\rm e}$ is the electron mass.
If the above condition does not hold,
too many positrons are reflected back and destroy the negative charge
region (cathode).
In order to screen the electric field consistently,
huge number of pairs should be injected within the small thickness 
$l_{\rm E}$.
The required pair multiplication factor per one primary particle
is enormously large and cannot be realized in the conventional pair
creation models. Thus, some other ingredients are required for
the electric field screening.

In the previous studies of the screening,
pairs were assumed to accelerate or decelerate
along the 0-th order trajectories determined by $E_\parallel$.
However, if an electrostatic (longitudinal) instability
occurs, the excited waves may produce effective friction.
Friction on particles change the charge polarization process.
Thus, instability in the presence 
of an external electric field may have a relevance to this problem, 
which motivates us to make an exploratory study in this paper.
Various instability mechanisms outside the accelerator have been studied 
for pair plasmas along with a primary beam in relation to 
coherent radio emission mechansims \citep{hin76, che77, ass80, ass83, lyu92}.
However, plasma instability inside the accelerator has not been studied.
The Lorentz factor of the primary beam ($\Gamma \simeq 10^6$--$10^7$)
is much larger than that of electron--positron pairs
($\gamma \simeq 10^2$--$10^3$).
In such a case, it is difficult to induce the two-stream instability.
However, it is not clear whether pairs stably flow
in the electric field $E_\parallel$ or not.
For typical pulsar parameters, the braking distance is
$l_{\rm E} \simeq 10^{-2}$ cm, while the length-scale
of plasma oscillation is 
$(m_{\rm e} c^2/4 \pi e^2 n_{\rm GJ})^{1/2} \simeq 1$ cm
\citep{shi98, shi02}, where $n_{\rm GJ}=|\rho_{\rm GJ}/e|$.
Particles are accelerated or decelerated in a period that is shorter
than the typical time scale of plasma oscillation.
Therefore, the distribution function is not uniform
on the scale we consider.
The local approximation is not adequate to deal with plasma oscillation.

Properties of a pair plasma in such a strong electric field have not been 
studied.
Such studies may bring us a new key to understanding astrophysical phenomena.
One of our purposes is to examine if electrostatic instability
makes the screening easier.
As the first step toward this purpose, in this paper we investigate
electrostatic instability of pairs injected
in an external electric field.
Investigations of electrostatic waves, when
we cannot adopt the local approximation, may be important
not only for pulsars but also for other high-energy astronomical phenomena. 
Since an analytical treatment is difficult in this case,
we simulate electrostatic waves numerically in idealized situations.
In \S 2 we mention the two-stream instability
without an electric field for comparison.
In \S 3 we describe the situation we consider.
The most simplified physical condition is adopted;
the electric field, injection rate, injection energy are constant.
In \S 4 we explain our numerical method.
Our method can treat only linear waves.
Numerical results are summarized in \S 5.
Our results show a new type of plasma instability
due to electric field.
\S 6 is devoted to summary and discussion.

\section{Two stream instability with local approximation}

First of all, for reference, we consider one-dimensional (1-D)
homogeneous flows of electrons and positrons in the absence of 
electric field. The pair-distributions are functions of the 
4-velocity $u=\beta/\sqrt{1-\beta^2}$, where $\beta=v/c$.
For simplicity, we assume that the distribution functions
are expressed by the step function $\Theta(u)$ as
\begin{eqnarray}
f_{-}&=&C \left[ \Theta(u-u_3)-\Theta(u-u_4) \right] \quad 
\mbox{for electrons,}
\label{fe}\\
f_{+}&=&C \left[ \Theta(u-u_1)-\Theta(u-u_2) \right] \quad 
\mbox{for positrons,}
\label{fp}
\end{eqnarray}
where $u_4>u_3 \geq u_2>u_1$ with constant $u_i$.
Electrons homogeneously distribute between $u_3$ and $u_4$ 
in $u$-space, while positrons similarly distribute between 
$u_1$ and $u_2$ (see Figure 1).
As will be seen in the next section, the distributions 
in an external electric field that we will consider
are similar to the above distribution functions.
In the linear perturbation theory,
the dispersion relation for electrostatic waves \citep{bal69} 
is given by
\begin{eqnarray}
D(k,\omega)=1+\sum_a \frac{4 \pi q_a^2}{\omega m_a}
\int d^3 u \frac{\beta}{\omega-k v} 
\frac{\partial f_a}{\partial u}=0,
\end{eqnarray}
where $q_a$ and $m_a$ denote charge and mass of the particle 
species $a$. Solutions of the dispersion relation usually 
yield a complex frequency.
The imaginary part of $\omega$ corresponds to the growth rate
of waves, $\omega_{\rm i}$.
A positive growth rate $\omega_{\rm i}>0$ implies an exponentially 
growing wave, while a negative $\omega_{\rm i}$ does an 
exponentially damped wave.
Adopting the distribution functions (\ref{fe}) and (\ref{fp}),
we obtain
\begin{eqnarray}
\tilde{\omega}+\beta_1/(\tilde{\omega}-\tilde{k} \beta_1)
-\beta_2/(\tilde{\omega}-\tilde{k} \beta_2)
+\beta_3/(\tilde{\omega}-\tilde{k} \beta_3) \nonumber \\
-\beta_4/(\tilde{\omega}-\tilde{k} \beta_4)=0,
\end{eqnarray}
where the subscripts $i$ denote corresponding values to $u_i$,
$\tilde{\omega}=\omega/\omega_{\rm p}$, 
$\tilde{k}=k c/\omega_{\rm p}$,
and $\omega_{\rm p}^2=4 \pi e^2 C/m_{\rm e}$.
The above equation is reduced to a quartic one.
Although it is messy to obtain the solutions analytically,
we can numerically confirm that $\tilde{\omega}$ has
complex values, for a given $\tilde{k}$.
The solutions obey the usual two-stream instability properties;
the larger the difference in velocities of electron and positron 
flows is, the smaller the maximum growth rate and the corresponding 
$\tilde{k}$ become.

When $u_2=u_3$, electrons and positrons distribute
continuously, though the average velocity is different.
In this case, the dispersion relation is reduced to a quadratic 
equation, and we obtain
\begin{eqnarray}
\tilde{\omega}=\tilde{k} (\beta_1+\beta_4)/2
\pm \frac{1}{2}\sqrt{\tilde{k}^2 (\beta_4-\beta_1)^2
+4 (\beta_4-\beta_1)}.
\end{eqnarray}
Apparently, $\tilde{\omega}$ is real.
Therefore, even though the velocities of the two flows are different,
instabilities are not excited in homogeneous pair plasmas,
as long as electrons and positrons distribute continuously in $u$-space.
This is because the absolute values of their charges and masses are the same for 
electron--positron pairs.
As will be shown in the next section,
the distribution of pairs injected in an external electric field
is similar to the above distribution locally.
However, we will show that an electric field induces
instability. 

\section{Pair injection in an external electric field}

In a strong magnetic field as in the pulsar polar caps,
transverse momenta of relativistic electrons and positrons
are lost during a very short time via synchrotron radiation.
These particles move along the magnetic field lines and 
their distribution functions are spatially 1-D.
In this section we consider 1-D distribution functions
of electron-positron pairs injected in an external electric 
field that is parallel to magnetic field lines.
As the conventional theories for the two-stream instability
implicitly assume, we neglect the toroidal magnetic field 
due to the global current of plasma.
Only within this treatment the 1-D approximation is adequate.
In order to simplify the situation,
we assume the external electric field $E_0$ is constant.
In pulsar models, there exists a primary beam which 
produces current flow.
The external electric field is determined
by the complicated combination of the beam current, injected pair plasma, and
GJ density.
The constant $E_0$ requires that the charge density of the beam, pairs, and GJ density
cancels out, which may be an artificial situation.
As will be discussed in \S 6, the approximation of constant $E_0$
is justified for a smaller rate of pair injection,
which may be realized for actual pulsar parameters.

Anyway, we depart from actual pulsar physics, and
deal with plasma physics in an idealized situation hereafter.
We adopt the 1-D approximation and assume the existence of
the background charge which leads to constant $E_0$.
In our treatment we totally neglects effects of the existence 
of the background
on development of waves, and consider the behaviour of pair plasma only.
Pairs are assumed to be injected between 
$z=0$ and $z_{\rm M}$ at a constant rate $\dot{n}_0$.
In our calculation the pair injection is monoenergetic with 
4-velocity $u=u_0>0$.
Let us start from assuming the steady state of flows of 
electron-positron pairs.
The distribution function $f_0(z,u)$ 
satisfies the Boltzmann-Vlasov equation
\begin{eqnarray}
v \frac{\partial f_0}{\partial z}+\frac{q E_0}{m_{\rm e} c}
\frac{\partial f_0}{\partial u}=\dot{n}_0 \delta(u-u_0),
\label{bol}
\end{eqnarray}
for $0<z<z_{\rm M}$.
Hereafter we assume $E_0<0$ for definiteness.

Then, injected positrons ($q=e>0$) will
be decelerated as
\begin{eqnarray}
u=u_0-(t-t_{\rm inj})/t_{\rm E},
\end{eqnarray}
where $t_{\rm E} = m_{\rm e} c/|e E_0|$, and $t_{\rm inj}$ is 
the injection time.
The value of $t_{\rm E}$ represents the time-scale in which
the Lorentz factor of decelerating particles decreases by one.
Positrons will be turned back following the trajectory
\begin{eqnarray}
z/l_{\rm E}=\gamma_0-\gamma+z_{\rm inj}/l_{\rm E},
\end{eqnarray}
where $\gamma =\sqrt{\mathstrut u^2+1}$, $\gamma_0 =\sqrt{u_0^2+1}$,
$l_{\rm E}=c t_{\rm E}$, and $z_{\rm inj}$ is the injection point.

Electrons ($q=-e<0$) will continue to be accelerated as
\begin{eqnarray}
u&=&u_0+(t-t_{\rm inj})/t_{\rm E}, \\
z/l_{\rm E}&=&\gamma-\gamma_0+z_{\rm inj}/l_{\rm E}.
\end{eqnarray}

In our model electrons and positrons distribute in
the 2-D phase space (spatially 1-D) as is illustrated in Figure 2.
In the phase space
the trajectory of electrons injected at $z=0$
provides the maximum 4-velocity $u_{\rm M}=u_{\rm M}(z)$ at $z$,
while $u$ of positrons injected at $z=z_{\rm M}$
corresponds to the minimum 4-velocity $u_{\rm m}=u_{\rm m}(z)$ at $z$.
The region enclosed by the curves in Figure 2
is composed from the family of trajectories of pairs.
As long as pairs are uniformly injected at a constant rate,
the distribution function is constant because of the Liouville's theorem.
The distribution functions
between $z=z_{\rm ret}=(\gamma_0-1)l_{\rm E}$ (see Figure 2) 
and $z_{\rm M}$
are the same as in Figure 1 with $u_1=u_{\rm m}$, $u_4=u_{\rm M}$,
and $u_2=u_3=u_0$.
In this region the distribution functions
are expressed as
\begin{eqnarray}
f_{0-}&=&n_0 \left[ \Theta(u-u_0)-\Theta(u-u_{\rm M}(z)) \right]
\quad \mbox{for electrons,}\\
f_{0+}&=&n_0 \left[ \Theta(u-u_{\rm m}(z))-\Theta(u-u_0) \right] \quad 
\mbox{for positrons,}
\end{eqnarray}
where $n_0=t_{\rm E} \dot{n}_0$.
It is easily confirmed that $f_{0-}$ and $f_{0+}$ satisfy
the Boltzmann-Vlasov equation (\ref{bol}).
Electrons and positrons distribute continuously.
The charge density given by Eqs. (11) and (12) is not strictly 
zero so that our approximation of constant electric field is 
not self-consistent, unless the background charge cancels the total charge.
If the background does not play such a role,
our treatment is correct only when $n_0$ is small enough.
The quantitative condition for $n_0$ will be discussed in \S 6.
As was discussed in \S 2,
if we neglect the electric field and adopt the local approximation,
no wave instability is generated for this continuous distribution .
However, in our consideration
the time-scale $t_{\rm E}$ is too short 
(or $l_{\rm E}$ is too short) to adopt the local approximation.

In the region $0<z<z_{\rm ret}$ the pair distribution has separate two streams.
Since this region is peculiar in our idealized model,
we do not consider the waves in this region hereafter.

\section{Numerical method}

We consider linear perturbations of the distribution function
and electric field as
\begin{eqnarray}
f&=&f_0(z,u)+f_1(z,u,t), \\
E_z&=&E_0+E_1(z,t),
\end{eqnarray}
where $|f_1| \ll f_0$ and $|E_1| \ll |E_0|$.
Since the unperturbed distributions of pairs are inhomogeneous,
we cannot carry out a Fourier-harmonic expansion
of the perturbations.
Therefore, we solve time development of the perturbations rather 
than obtain the linear modes. 
As we have seen in the previous section,
our idealized situations lead to the simple distribution function $f_0$,
which makes numerical computaions easier. 

We directly solve the perturbations $f_1$ and $E_1$
from the linearly perturbed Boltzmann-Vlasov equation
\begin{eqnarray}
\frac{D f_1}{D t} \equiv
\frac{\partial f_1}{\partial t}+v \frac{\partial f_1}{\partial z}
+\frac{q E_0}{m_{\rm e} c} \frac{\partial f_1}{\partial u}=
-\frac{q E_1}{m_{\rm e} c} \frac{\partial f_0}{\partial u},
\label{bol2}
\end{eqnarray}
where $D t$ is the differential along the 0-th order trajectory.
The Amp\`ere-Maxwell law is written by 
\begin{eqnarray}
\frac{\partial E_1}{\partial t}=-4 \pi j,
\end{eqnarray}
where $j$ is the perturbed current density.
Initially $E_1$ depends on only $z$, and the magnetic field 
$B_x=B_y=0$.
Then, the Faraday law ensures that $B_x$ and $B_y$ remain 
zero all the time.
The component $B_z$ does not affect time developments of $E_z$
and $f_1$.

We set up grids along the 0-th order trajectories of pairs 
in the 2-D phase space $(s =z/l_{\rm E},u)$.
Following the Lagrangian method
we follow time evolution of $f_1$ in these grids from equation (\ref{bol2}).
For $s_{\rm ret} \leq s \leq s_M$, equation (\ref{bol2}) is rewritten as
\begin{eqnarray}
\frac{D F_-}{D \tau}&=&{\cal E}(s) 
\left[ \delta(u-u_0)-\delta(u-u_{\rm M}(s)) \right], 
\label{f-} \\
\frac{D F_+}{D \tau}&=&-{\cal E}(s) 
\left[ \delta(u-u_{\rm m}(s))-\delta(u-u_0) \right],
\label{f+}
\end{eqnarray}
where we introduce dimensionless values
$F=f_1 l_{\rm E}$, $\tau=t/t_{\rm E}$, and 
${\cal E}=l_{\rm E} n_0 E_1/|E_0|$.
At the injection of pairs ($u=u_0$) $F$ obtains a finite value 
in proportion to the perturbed electric field ${\cal E}$.
Then, the value of $F$ is propagated along the 0-th order trajectory  
(see Figure 2), i.e., $F$ is conserved along the characteristics. 
The disturbance  $F_+$ initially propagates forward,
and then turns back at a distance $\gamma_0-1$ from the injection point,
while $F_-$ simply propagates forward.
For $|u|<u_0$, $F_+(u,s)$ is originated from inner injection positions 
between $s-\gamma_0+1$ and $s$.
On the other hand, for $|u|>u_0$, $F_+(u,s)$ is originated from 
the outer injection positions $>s$.
The values of $F$ on the trajectories of the pairs injected at $z=0$ 
and $z=z_{\rm M}$
are special and they will continue to be changed by the electric field
during propagation.

On the other hand, the evolution of electric field is calculated by 
the Eulerian method;
\begin{equation}
\frac{\partial {\cal E}}{\partial \tau}= R^2 \Biggl[ \int du \beta_- F_-
-\int du \beta_+ F_+ \Biggr],
\label{AM}
\end{equation}
where $R=\omega_p t_{\rm E}$ with the plasma frequency
$\omega_p=\sqrt{4 \pi e^2 n_0/m_{\rm e}}$.
The parameter $R$ is the ratio of the braking time-scale
to the typical time-scale of the plasma oscillation.
In our case $R \propto t_{\rm E}^{3/2} \propto |E_0|^{-3/2}$.
In the pulsar polar cap model electric field
is so strong that $R$ is much smaller than one.
Since the charge conservation is assured by the Boltzmann-Vlasov equation,
the Gauss law,
\begin{eqnarray}
\frac{\partial {\cal E}}{\partial s}= R^2 \Biggl[ \int du F_+
-\int du F_- \Biggr],
\end{eqnarray}
is automatically satisfied, if the law is initially satisfied.
The configuration of ${\cal E}$ is determined by the charge density
that is an integral of $F_+ - F_-$.
Since the propagations of $F_+$ and $F_-$ are complex,
it is difficult to predict the behaviour of ${\cal E}$ intuitively.

We have ascertained that results obtained from
our numerical code satisfy the Gauss law.
In addition we have checked our code by reproducing
two stream instability in the absence of electric field,
using the distribution functions in \S 2.

\section{Results}

We have simulated electrostatic waves from various initial conditions 
and parameter values.
We are interested in a parameter region $R <1$.
In this region the typical wavelength of plasma oscillation 
$\sim l_{\rm p}=c/\omega_p$
is longer than the braking distance $l_{\rm E}=R l_{\rm p}$.
We give an initial disturbance in a spatially limited region. 
As will be shown below, when $R u_0 \geq \sim 1$, 
we find an absolute instability
in which disturbance grows in amplitude but always embraces the 
original region, where the initial disturbances of $F$ and 
${\cal E}$ are given.
The condition $R u_0 \geq 1$ means that the distance 
injected positrons move forward 
before they turn back, $l_{\rm E} (\gamma_0-1) \sim l_{\rm E} u_0$, 
is larger than the length-scale of the plasma oscillation $l_{\rm p}$.
On the other hand, for $\sim 0.1 \leq R u_0 \leq \sim 1$,
we find a convective instability in which
disturbance grows while propagating away from the original region.
The waves excited from the convective instability propagate backward.
Empirically, the results do not largely depend on
the spatial size of the pair injection region $s_M=z_{\rm M}/l_{\rm E}$,
which determines the minimum 4-velocity $u_{\rm m}(z)$.
In this section we show some examples of the instabilities found 
in our simulations.
The parameters and initial conditions are summarized in Table 1.

The initial conditions are taken to satisfy the Gauss law.
Given the parameters $R$, $u_0$, and $s_M$,
we set the initial values of the disturbances $F$ and ${\cal E}$ as
\begin{eqnarray}
{\cal E}(s) &=& {\cal E}_{\rm i}
( \cos{( k_{\rm i} (s-s_{\rm i} ) )}-1 ), \\
F_-(s,u) &=& \frac{(k_{\rm i}-k_\Delta) {\cal E}_{\rm i}}
{2 R^2} \sin{(k_{\rm i} (s-s_{\rm i}) ) } \delta(u-u_0), \\
F_+(s,u) &=& -\frac{(k_{\rm i}+k_\Delta) {\cal E}_{\rm i}}
{2 R^2} \sin{(k_{\rm i} (s-s_{\rm i}) ) } \delta(u-u_0),
\end{eqnarray}
for $s_{\rm i} \leq s \leq s_{\rm i}+2 \pi/k_{\rm i}$.

The initial disturbances are confined within a small
spatial region of one wavelength ($2\pi/k_{\rm i}\ll s_{\rm M}$).
In the other region there is no disturbance.
The initial perturbation of ${\cal E}(s)$
has a single sign with a form of a cosine curve.
On the other hand, $F_-$ and $F_+$ have the form of a sine curve, 
satisfying the Gauss law.
The parameter $k_\Delta$ induces asymmetry of the charge density
of electrons and positrons.
The total charge density ($\propto F_+ - F_-$)
does not depend on $k_\Delta$, and also has a
form of a sine curve with $k_{\rm i}$.
We have tried various values of $k_\Delta$ in our simulations and 
find that the ratio of $F_-$ to $F_+$ is settled as time passes 
irrespective of $k_\Delta$.
When instabilities occur, growing wave modes end up 
dominating other modes.
Therefore, results do not largely depend on the initial conditions.

First we describe the results for $R=0.1$ (the calculations RUN1--RUN3) 
and see the behaviour of the linear perturbations for various 
values of the parameter $u_0$. 
In Figures 3 and 4 we plot electric field ${\cal E}$
for RUN1 for which $R u_0=1$.
In this calculation, the initial disturbance exists from 
$s_{\rm i}=120$ to $s_{\rm i}+2 \pi/k_{\rm i} \simeq 180$.
Since positrons will turn the direction of motion after 
$\tau\sim 10$, we must follow the disturbance much longer than that. 
As is illustrated in Figure 3, at $\tau=25$ the disturbance
${\cal E}$ remains in the originally disturbed region.
As time passes, the amplitude around the original region of the disturbance
grows, and the wave packet spreads backward little by little.
In the forward region $s> \sim 200$, we do not observe any growing wave.
Although particles move almost at the light velocity,
the disturbances remain around the original region
and the wave packet does not spread at the light velocity.
In order to show the growth of the amplitude, in Figure 5
we plot the time evolution of $E_{\rm M}$ that is the electric field
for the maximum amplitude. 
The maximum electic field $E_{\rm M}$ oscillates over
positive and negative regions.
Initially $E_{\rm M}$ changes complicatedly because of
the initial conditions we artificially set.
As time passes, $E_{\rm M}$ smoothly grows while oscillating.
The period of the oscillation of $E_{\rm M}$
is $\simeq 20 t_{\rm E}=2/\omega_p$.
The growing time $t_i$, where $|E_{\rm M}| \sim \exp{(t/t_i)}$, is
$\simeq 50 t_{\rm E}=5/\omega_p$.

Let us look into the the behaviour on a shorter time scale for RUN1.
At a fixed position $s$, the local electric field ${\cal E}(s)$ grows 
while oscillating. However, when we see spatio-temporal behaviour, 
we notice that the spatial pattern propagates backward
while changing their amplitude.
As is shown in Figure 4, waves propagate from $s \simeq 200
\simeq s_{\rm i}+2 \pi/k_{\rm i}$ backward with a growing amplitude.
The amplitude becomes maximum around $s_{\rm i}=120$,
and then the amplitude declines while propagating backward.
This decline leads to the confinement of the wave packet.
Even though waves pass the disturbed region many times,
waves exist only in a spatially limited region.
In the wave packet of ${\cal E}(s)$ there are multiple peaks and bottoms,
and the most prominent one of them corresponds to $E_{\rm M}$.
If we define the `phase velocity' as the velocity of peaks (or bottoms),
the phase velocity ($\simeq -2.8 c$) turns out 
to be faster than the velocity of light.
As peaks propagate backward,
the peak or bottom associated with $E_{\rm M}$ alternates one after another,
so that the position of $E_{\rm M}$ hangs around the original region.
When we define the group velocity by averaging the velocity of the 
position of ${\cal E}=E_{\rm M}$ for a longer time scale than
the oscillation period, the group velocity turns out 
to be almost zero.
The wavelength $\lambda$ is about $60 l_{\rm E}\simeq 2 \pi l_{\rm p}$
which is almost the same as the initial wavelength
of the disturbance.
Even if we start from another $k_{\rm i}$,
the growing mode dominates others and the final wavelength is the same
as this result, $60 l_{\rm E}$.
The final wavelength of growing waves is unchanged for different
initial conditions.
Figure 6 shows charge density distributions at $\tau=195$.
The number density of electrons (positrons) has opposite (same) sign of
the charge density in Figure 6.
The phases of number densities of electrons and positrons
are the same.
The amplitude of the positron density is always larger than
that of the electron density.
The difference in the number densities of positrons and electrons is proportional to
the total charge density which satisfies the Gauss law.

Next we discuss the results of RUN2 ($R u_0=30>1$). 
The initial distubance ranges from 
$s_{\rm i}=500$ to $s_{\rm i}+2 \pi/k_{\rm i} \simeq 560$.
In Figure 7 we show electric fields at several epochs.
The wave profiles are not so simple compared to the case of RUN1
($R u_0=1$).
There are waves propagating both forward and backward.
This may be because the distance positrons move forward 
before they returns
($\sim 300 l_{\rm E}$
from their injection point) is longer than
the typical wavelength $2 \pi l_{\rm p} \simeq 60 l_{\rm E}$.
Though the properties of the waves are complex,
we can see that there is an instability in this case, too.
The waves around the original region grow while diffusing 
both forward and backward.
We note that a separate component of the disturbance appears 
around $z < z_{\rm ret}$ ($s<300$ in this case). This disturbance
may be due to two stream instability,
which grows faster than in the other region.

Although we do not show here, for a much larger value of $R u_0$, 
absolute instabilities are found to occur in our simulation.
However, we do not further pursue this issue 
because we need to calculate over a much wider region of $s$ and 
resultant memory in computation becomes large for a large value 
of $u_0$.

In cases for $R u_0=0.3<1$ (the calculations RUN3), absolute instability
does not occur, but convective instability propagating backward occurs 
(see Figure 8).
In RUN3 the initial disturbance is given from $s=420$ to $\sim 426$
with a wavelength $2 \pi l_{\rm E}/k_{\rm i} \simeq 6 l_{\rm E}$.
The disturbance of the electric field is seen to propagate backward,
and before long characteristic waves grow.
The wavelength of the growing wave ($\simeq 9 l_{\rm E}$) is slightly
longer than the initial length and almost constant.
As the disturbance propagates,
the wave packet spreads and the number of waves in the packet
increases.
The growing time of $E_{\rm M}$ is about $100 t_{\rm E}=10/\omega_p$
(see Figure 9).
The phase velocity, $v_{\rm ph} \simeq -1.09 c$,
is faster than the velocity of light.
The amplitude of each peak in the packet initially grows.
As the peak approaches the head of the wave packet,
the growth of the amplitude turns over to damping.
This behaviour is similar to that in the absolute instability for RUN1.
We obtain the group velocity $v_{\rm g} \simeq -0.85 c$.
Figure 10 shows that the charge density is dominated by positrons.
We have simulated for $R u_0=0.3$ with various initial conditions.
However, in any case there is no sign of wave instability propagating 
forward.

We have investigated for $R=0.01$ also (the calculations RUN4--RUN6),
and confirmed that the qualitative results
are determined by the value of $R u_0$ (see Table 2).
The absolute instability and convective instability are
induced for $R u_0=1$ and $R u_0=0.3$, respectively.
For $R u_0=0.03$ (RUN6), however, we do not find any instability.
A smaller value of $R$ means lower density of pair plasma for 
a given value of the external electric field.
In such a low density plasma, particles tend to be less affected 
by forces from other particles compared to the external electric field.
Therefore, too small value of $R \ll 1/u_0$ makes the plasma stable.
From the above results, we may conclude that
electrostatic instability is mainly determined by the parameter $R u_0$.
The wavelength and the growing time are a few or ten times
the typical scales of the plasma oscillation, $l_{\rm p}$ and
$1/\omega_{\rm p}$, respectively.

The value $R$ has been assumed to be smaller than 1 heretofore.
We have also simulated for $R \geq 1$ and found instabilities.
The wave properties are as complicated as the example in RUN2,
so that we do not report the details of the results in this paper.
When $R \geq 1$,
$E_0$ is so small that $l_{\rm E} \geq l_{\rm p}$.
In the limit of $E_0=0$ ($R \to \infty$),
the instability does not occur as was shown in \S 2.
However, we  have not tried the cases $R \gg 1$,
because of poor computational capacity so far.

Judging from the backward spread of wave packets and the negative
phase velocity, it is seen that returning positrons play a decisive
role in the instabilities.
The dominance of positrons in the charge density
in comparison with electrons also suggests that
the instabilities are due to returning positrons.
It is remarkable that
positrons pass the same region twice, forward and backward.
The typical wavelength of plasma oscillation ($2 \pi l_{\rm p}$)
may resonate with the distance positrons move forward,
$(\gamma_0-1) l_{\rm E}$.
As we have mentioned in \S 4,
the excited electric field ${\cal E}$ generates
the disturbances of pairs, $F(u_0)$, at their injection.
The value of $F(u)$ is transported along the trajectories of pairs,
conserving their value.
Since $F_+$ and $F_-$ acquire the same value at their injection,
the contribution to the charge density is almost canceled out 
as long as electrons and positrons
move forward together at $\sim c$.
When positrons turn around, the charge is polarized.
As for positrons, $F_+(u,s)$ at $\tau$
conserves the information on the electric field ${\cal E}(s')$
in the past ($\tau'=\tau-(u_0-u)$), where 
$s=s'+\gamma_0-\gamma$.
The displacement between $s$ and $s'$ conforms to the 
trajectories of positrons.
Therefore, the charge density of positrons (except for $u=u_{\rm m}$)
is proportional to a superposition of displaced ${\cal E}(s')$ 
at different times.
This superposition will increase the amplitude of the charge density
in response to evolution of ${\cal E}$.

Since $F_+(u \neq u_{\rm m})$ is constant along the characteristics, 
the value $F_+$ remains to be finite even after positrons enter 
the region where ${\cal E} \sim 0$,
while $F_+(u_{\rm m})$ changes as long as ${\cal E}$ exists.
Our numerical results imply that $F_+(u_{\rm m})$ 
cancels out the charge density
due to $F_+(u \neq u_{\rm m})$ in the region where ${\cal E} \sim 0$.
This process prevents the waves from spreading backward at the speed
of light.
The propagation of the disturbance by electrons
is relatively simple.
The contribution due to $F_-(u_{\rm M})$
prevents the waves from spreading forward,
as $F_+(u_{\rm m})$ does.
In RUN3 and RUN5 ($R u_0 <1$), positrons turn back and move backward quickly
before $F_+(u_{\rm m})$ cancels out the charge density,
so that growing waves propagate backward.

\section{Conclusions and Discussion}

In this paper we have found a new type of instability
in electron-positron pair flows injected in an external electric field,
which is assumed to be spatially constant.
The properties of the instability
are characterized by the ratio $R$ (the braking time-scale
to the typical oscillation time-scale of the plasma $1/\omega_{\rm p}$) and
4-velocity $u_0$ at injection.
For $R u_0> \sim 1$ absolute instability is induced,
while convective instability propagating backward is excited
for $\sim 0.1 < R u_0< \sim 1$.
The growing time in amplitude is as short
as a few times the time-scale $1/\omega_{\rm p}$.
The wavelength is also several times $l_{\rm p} =c/\omega_{\rm p}$.
The instabilities are caused by returning positrons.
For $R u_0 \ll 1$, the pair plasma turns out to be stable.
A small value of $R$ implies that the plasma density
is so low compared to the electric field $E_0$.
For $R \ll 1/u_0$ the collective interaction of the pair plasma
is not important, so that each particle moves along
the trajectory determined by $E_0$ independently of other particles.

Growing electrostatic waves may work as frictional forces.
In this paper we have treated waves as linear perturbations,
following the propagation of disturbances in the distribution function 
and electric field.
Our method does not allow us to follow processes
of gaining or losing kinetic energy of each particle from the waves.
The quasi-linear theory is not applied 
to deal with the reaction of particles as it is,
because the disturbances do not have the Fourier-harmonic dependence.
Thus, we consider the qualitative character of the effective reaction force
from a numerical treatment as follows.
The spatial averages of $F$ and ${\cal E}$ oscillate with time
in our simulations.
Therefore, the expectation values of $F$ and ${\cal E}$
can be considered to be zero.
On the other hand, when waves grow or are attenuated,
the spatial average of the cross term, $\langle F {\cal E} \rangle$,
may have a finite value.
As is the case with the quasi-linear theory,
the 0-th distribution function may change,
following the 2-nd order order approximations of the Boltzmann-Vlasov equation:
\begin{eqnarray}
\frac{\partial \langle f_0 \rangle}{\partial t}=-\langle \frac{q E_1}{m_{\rm e} c}
\frac{\partial f_1}{\partial u} \rangle
\propto - \frac{\partial G(u)}{\partial u},
\end{eqnarray}
where
\begin{eqnarray}
G(u) &=& \frac{q}{|q|}\int ds F(u,s) {\cal E}(s),
\end{eqnarray}
and $\langle f_0 \rangle =n_0$ independently of $u$ in our case.
In view of the Fokker-Planck approximation,
$G(u) \propto \langle \dot{u} \rangle f_0$,
where $\langle \dot{u} \rangle$ is the average change of $u$ due to the reaction force
per unit time.
Since $f_0$ is constant in our simulation,
$G(u)$ is proportional to the reaction force.

We plot $G(u)$ in Figure 11 for $\tau=195$ in RUN1.
The modulation pattern of $G(u)$ does not change,
but the amplitude grows with time.
Apparently, the modulation pattern is asymmetric
for particles of $u>u_0=10$ (electrons) and $u< u_0$ (positrons).
These qualitative behaviour is common for the other RUNs.
As Figure 2 and equations (\ref{f-}) and (\ref{f+}) show,
perturbations are generated at $u=u_0$
and $u=u_{\rm M}$ (or $u_{\rm m}$).
Figure 11 shows $G(u)$ for a region around $u=u_0$ only,
and outside of this region $G(u)$ has also significant value
due to the disturbances generated at $u=u_{\rm M}$ and $u=u_{\rm m}$.
However, the modulation pattern of $G(u)$ for such regions 
oscillates with time.

In the usual two-stream instability,
the excited waves accelerate background fluid,
and decelerate beam fluid.
As is shown in Figure 11,
the direction of the reaction force depends
on $u$ even in the same species of particles.
The amplitude of $G(u)$ takes always the maximum value at $u=u_0$.
If the effective reaction force grows enough,
positrons (electrons) just injected ($u \simeq u_0$) feel 
positive (negative) force
as is shown in Figure 11.
Thus, the reaction force may make particles tend to
stay around the regions of $u=u_0$.
The integral of $G(u)$ around $u=u_0$ (roughly from $u=-100$ to 
$10$ for Figure 11)
for positrons is also positive.
Such positrons are accelerated by the reaction force on average.
However, the integral becomes negative all the time,
if we include the contribution due to the disturbances
generated at $u=u_{\rm m}$,
though $G(u)$ for a large $|u|$ oscillates with times.
The returning positrons injected at $z=z_{\rm M}$ feel a negative
reaction force on average.
On the other hand, the absolute value of the integral for electrons
is much smaller than those for positrons.
Therefore, the reaction force does not work as
usual `frictional' force between electrons and positrons.
The particles just injected are most affected by the reaction force,
and lose (or gain) their energy owing to waves. 
If the reaction force is strong enough,
positrons just injected may have difficulty to turn back.
If positrons suffer from such frictional force,
the distribution function $f_0$ should be largely altered. 
This may help to solve the problem of the electric field screening 
in the pulsar polar caps.

If excited waves grow enough to change
trajectories of particles,
the waves cannot be treated as the linear perturbations.
In a strong electric field,
$|f_1|$ can be as large as $f_0$ before $|E_1| \sim |E_0|$ achieves.
We may simplify the energy-loss process of particles due to 
perturbed electric field as follows;
particles from $u=u_0$ to $u_0+\Delta u$
lose their energy owing to the waves at the same rate,
where $\Delta u$ is the equivalent width of particles interacting
the waves.
Then, the growth rate of the field energy is roughly considered as the average
energy loss rate of pairs,
which means $\omega_i E_1^2/4 \pi \sim |\dot{\gamma} \Delta u| n_0 m_{\rm e} c^2$,
where $\dot{\gamma}$ is the temporal change of $\gamma$ due to
the reaction force.
Here, we assume that the energy density of the waves
attributed to induced particle motions
is comparable or negligible to
the energy density of the electric field $E_1$.
When $|\dot{\gamma}| > 1/t_{\rm E}$, the frictional force
is sufficient to alter trajectories of pairs.
The above condition is rewritten as
$\omega_i E_1^2/\omega_{\rm p}^2 >m_{\rm e} c |E_0| \Delta u/e$.
Assuming $\omega_i \simeq \omega_{\rm p} \simeq 10^8$, $\Delta u=50$,
and $|E_0| \simeq 10^5$ in esu (these values may be typical for the pulsar polar cap),
$|E_1|$ is needed to be larger than $5 \times 10^3$ in esu to
change the distribution of pairs.
Even if $|E_1| < |E_0|$, the excited disturbances may change
the 0-th distribution of pairs.
However,
we need numerical simulations, which can deal with non-linear process,
in order to check how the reaction force modifies the distribution function $f_0$.

As a first step to deal with behaviour of pairs in an electric field,
in this paper we have assumed that the background charge distribution
cancels out the modification of $E_0$ due to injected electron-positron pairs.
Of course, this simplification may not be appropriate for pulsars,
while it makes computation easier.
Inhomogeneous electric field might play an important role
in plasma instability.
Let us check whether our model can be used when the background charge density
is constant for $s>s_0$.
The charge density changes for $s>s_0$ owing to the pair injection
and electric field should be modified.
The charge density decreases with distance
as $\propto \sim n_0 s$ for $s>s_0$.
In this case the variation of $\Delta E_0$
over the moving distance of injected positrons
before they turn back $l_{\rm E} u_0$ is $\sim (R u_0)^2 E_0$.
Therefore, in the case of stable plasma ($R u_0 \ll 1$)
the constant electric field is a good approximation
even for the constant background.
Although our simulations show a new possibility
of plasma instability around pulsars,
we need to simulate with an inhomogeneous
electric field for $R u_0 \sim 1$ in order to conclude whether
instability occurs in actual situations on pulsars.
In any case, the condition $R u_0 > \sim 0.1$ is an necessary
(but not sufficient so far) condition to induce the instability.

Let us consider implications for the pulsar polar cap.
We suppose that the primary electron-beam is accelerated from $z=0$,
and its Lorentz factor becomes $\Gamma$ at the pair production front (PPF) ($z=L$).
In this case the average electric field is $|E_0|=\Gamma m_{\rm e} c^2/e L$.
The braking time is expressed as $t_{\rm E}=m_{\rm e} c/|e E_0|=L/c \Gamma$.
By curvature radiation an electron of $\Gamma$ emits
$2 e^2 \Gamma/3 r \hbar \equiv M$ photons per unit time,
where $r$ is the radius of curvature of the field line.
We express the number density of the primary beam as $n_{\rm b}=f_{\rm b} n_{\rm GJ}$.
Assuming curvature gamma-rays immediately turn into pairs,
the pair-injection rate is approximated as $\dot{n}_0 \simeq M f_{\rm b} n_{\rm GJ}$.
Adopting the average electric field, the ratio $R$ becomes
\begin{eqnarray}
R&=&\omega_{\rm p} t_{\rm E}=\sqrt{\frac{4 f_{\rm b} e^3 \Omega B L^3}{3 m_{\rm e} c^4
\hbar r \Gamma^2}} \\
&\simeq& 2 \times 10^{-5} f_{\rm b}^{1/2} T_{0.3}^{-1/2}
B_{12}^{1/2} L_4^{3/2} r_7^{-1/2} \Gamma_6^{-1},
\end{eqnarray}
where $\Gamma_6=\Gamma/10^6$, and $T_{0.3}$, $B_{12}$, $L_4$,
and $r_7$ are in units of 0.3 s, $10^{12}$ G, $10^4$ cm,
and $10^7$ cm, respectively.
The average Lorentz factor of
pairs is at most $\hbar \Gamma^3/(2 m_{\rm e} c r) \simeq 2 \Gamma_6^3 r_7^{-1}$.
Therefore, we obtain
\begin{eqnarray}
R u_0 \simeq 4 \times 10^{-5} f_{\rm b}^{1/2} T_{0.3}^{-1/2}
B_{12}^{1/2} L_4^{3/2} r_7^{-3/2} \Gamma_6^{2}.
\end{eqnarray}
Even for the large values of $\Gamma$ and $L$ we can suppose in pulsar models,
we may expect $10^{-4} < R u_0 < 10^{-2}$ at most.
The electric field in the polar cap may be too strong to induce
the instability, compared to the pair-plasma density,
while the approximation of constant $E_0$ may be not so wrong in such cases.
If the electric field is much smaller than the average one $\Gamma m_{\rm e} c^2/e L$
at PPF as in the model in Aron's group \citep{faw77, sch78, aro79},
$R u_0$ can be large enough to induce the instability.
In such models, however, the pair polarization is not so important
to achieve the screening, while there is possibility
that the instability affect the radio emission process.

At present it is not clear that the electrostatic instability
we have considered in this paper is an important process
for the screening of electric field above the pulsar polar cap.
However, there may be extreme environments (magnetars etc.), where
$R u_0$ is as large as one.
We expect that studies on the plasma instability in electric fields
lead to opening a new approach to high-energy astrophysics.

\begin{acknowledgements}

This work is  supported in part by a Grant-in-Aid for Scientific Research 
from Ministry of Education and Science (No.13440061, F.T.).
One of the authors (K.A.) is supported by the Japan Society for the Promotion of Science.

\end{acknowledgements}

\clearpage

\begin{figure}
   \centering
   \includegraphics[width=\textwidth]{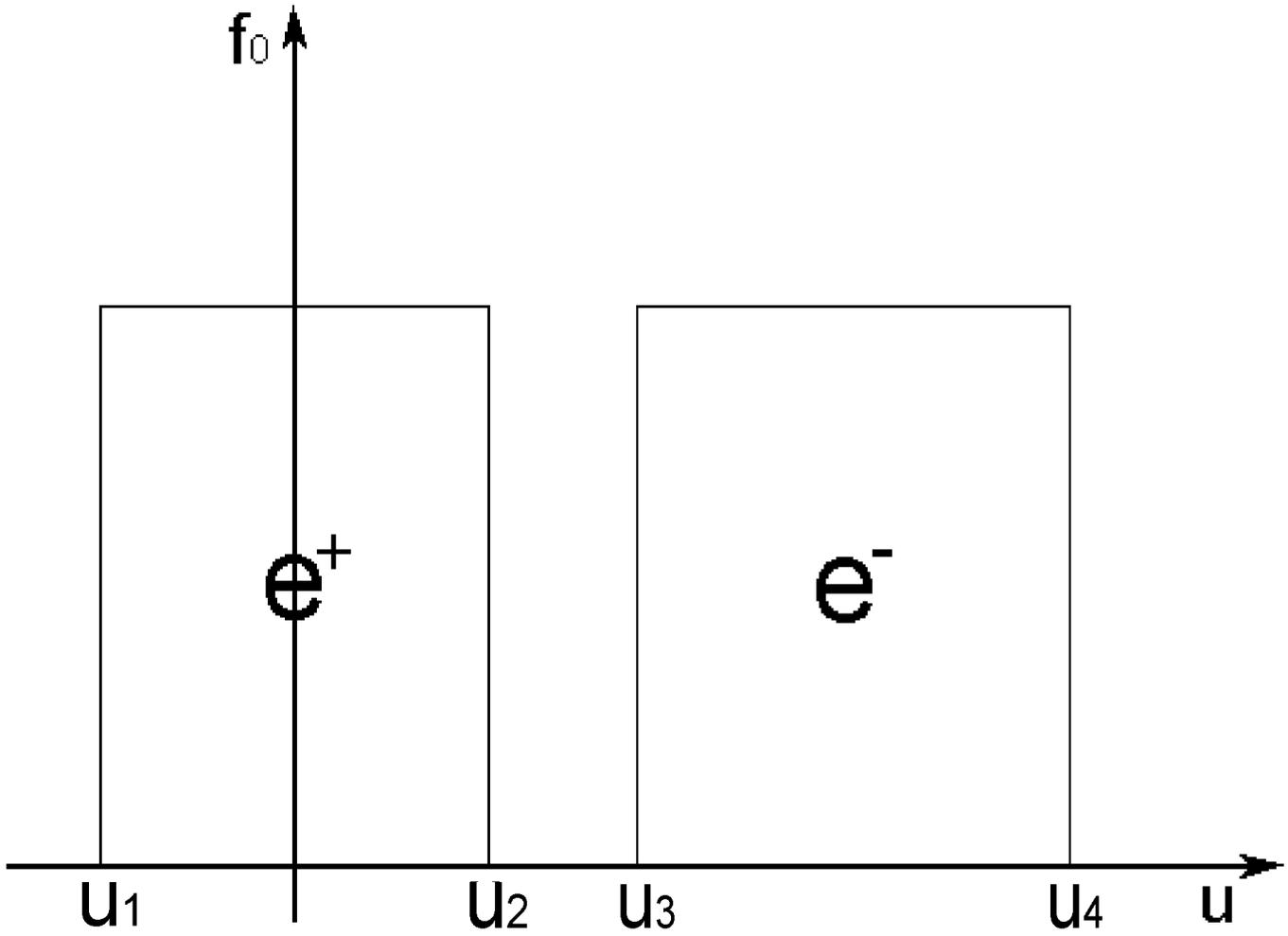}
\caption{Distribution function of the toy model in \S 2.}
\end{figure}

\clearpage

\begin{figure}
   \centering
   \includegraphics[width=\textwidth]{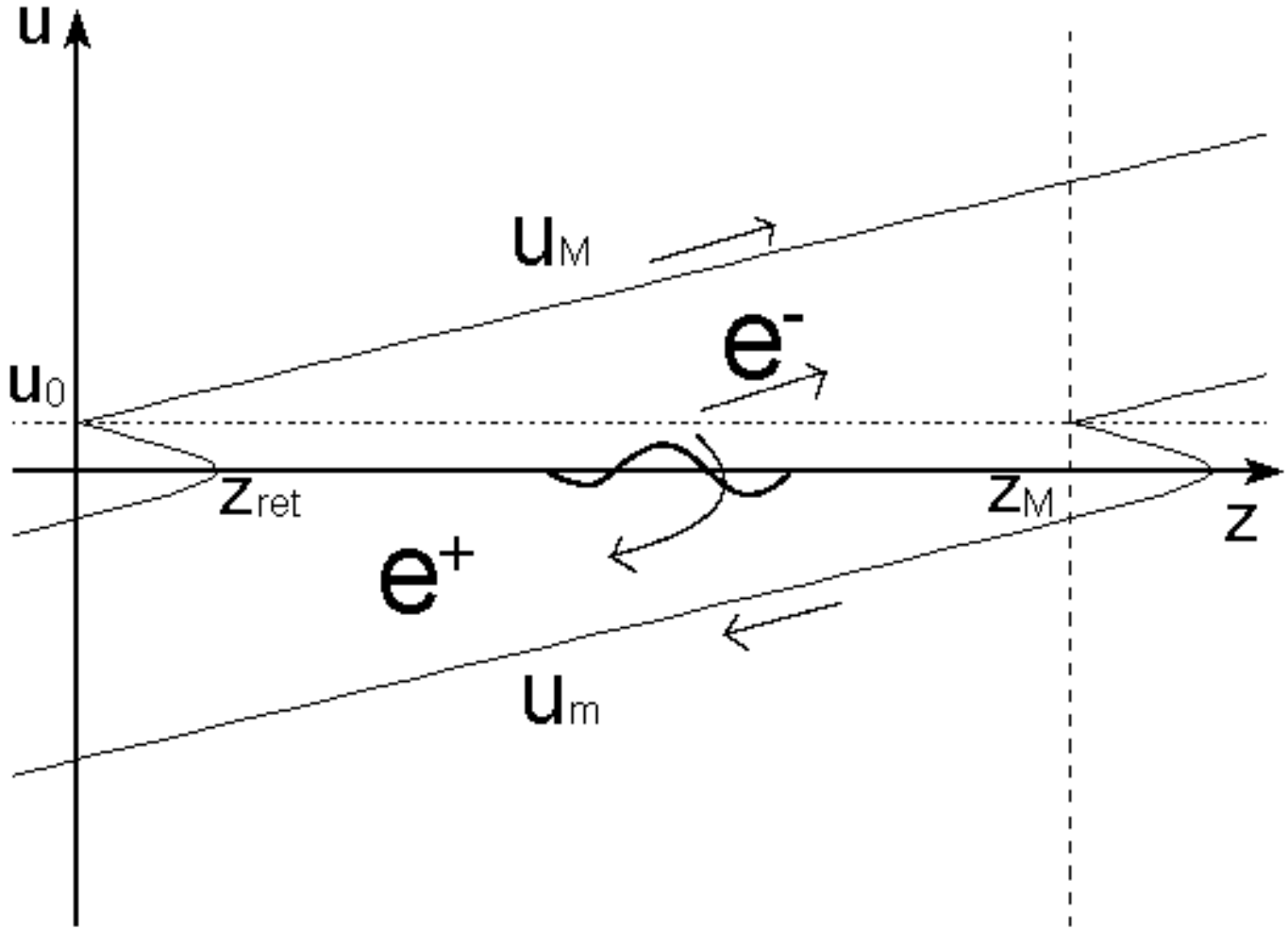}
\caption{Distributions of electrons and positrons in the phase space.
Electrons (positrons) are in the region $u \geq u_0$ ($u \leq u_0$).
The electric field ${\cal E}$
(schematically shown as the central corrugation in the figure) generates
the disturbances $F$ at the injection of pairs ($u=u_0$),
$u=u_{\rm M}$, and $u=u_{\rm m}$.
The arrows schematically show the transports of $F$ (see text).
}
\end{figure}

\clearpage

\begin{figure}
   \centering
   \includegraphics[width=\textwidth]{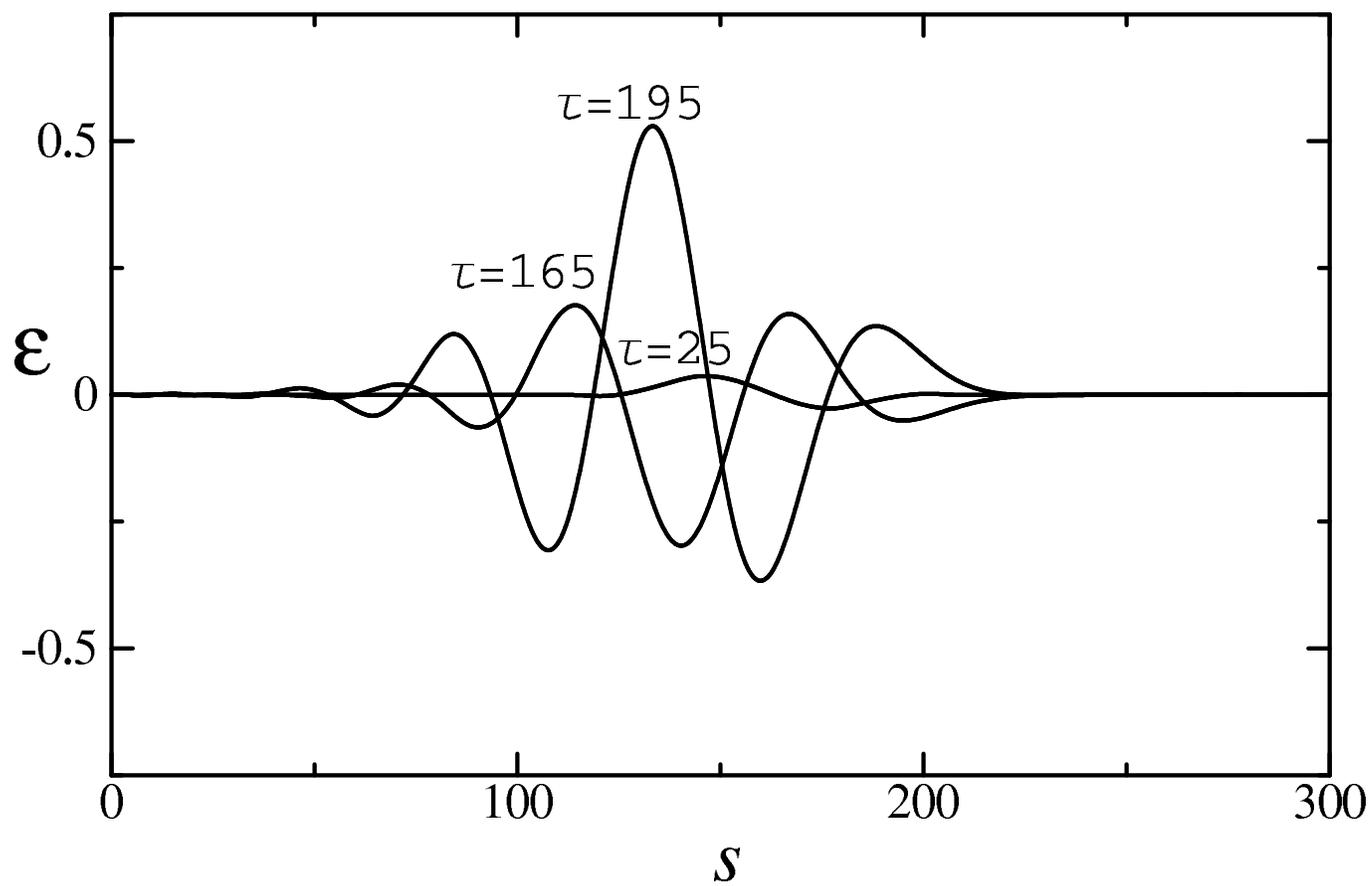}
\caption{Electrostatic waves for RUN1 for $\tau=25$, 165, and 195.
}
\end{figure}

\clearpage

\begin{figure}
   \centering
   \includegraphics[width=\textwidth]{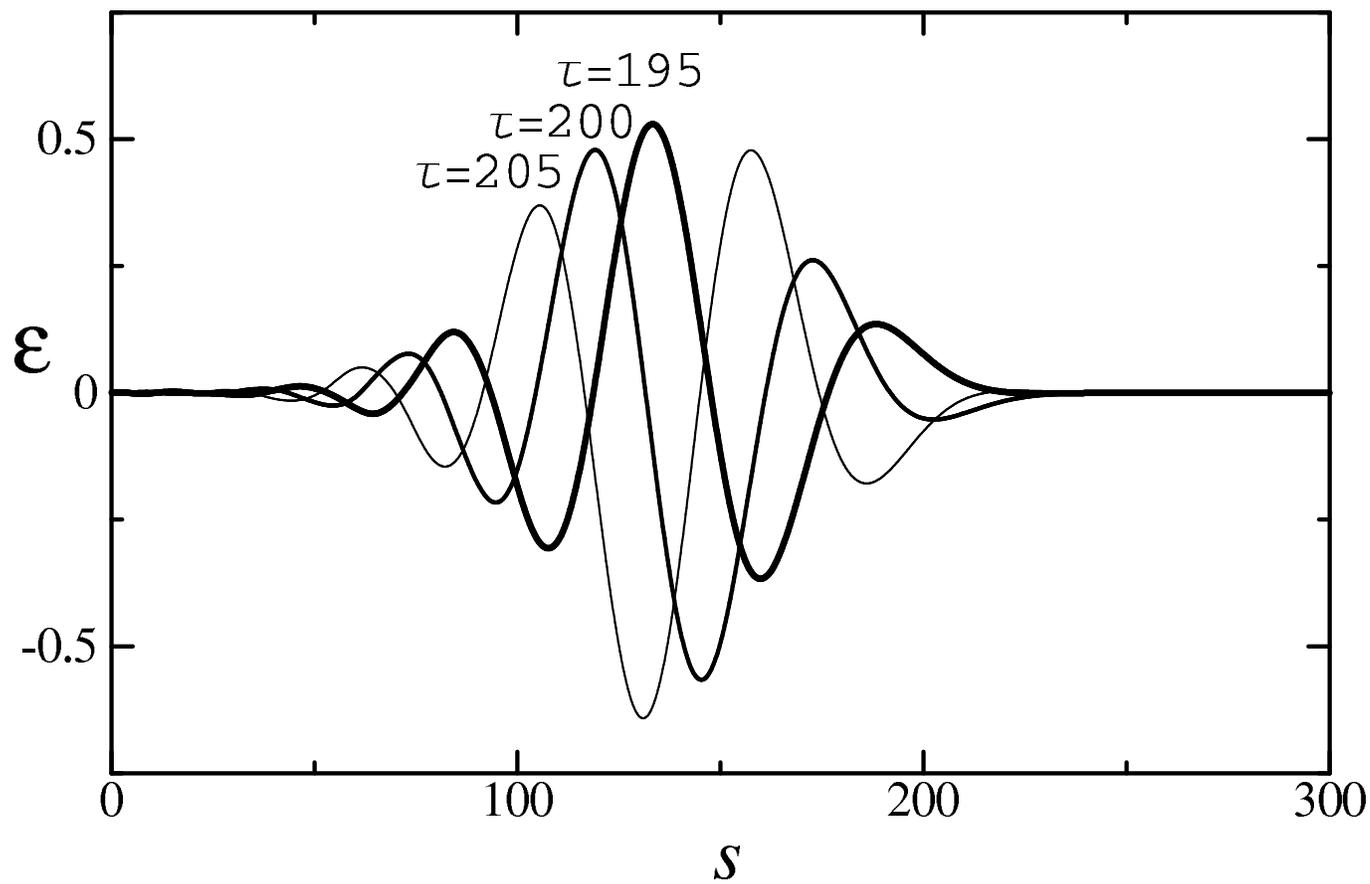}
\caption{Electrostatic waves for RUN1 around $\tau=200$.
}
\end{figure}

\clearpage

\begin{figure}
   \centering
   \includegraphics[width=\textwidth]{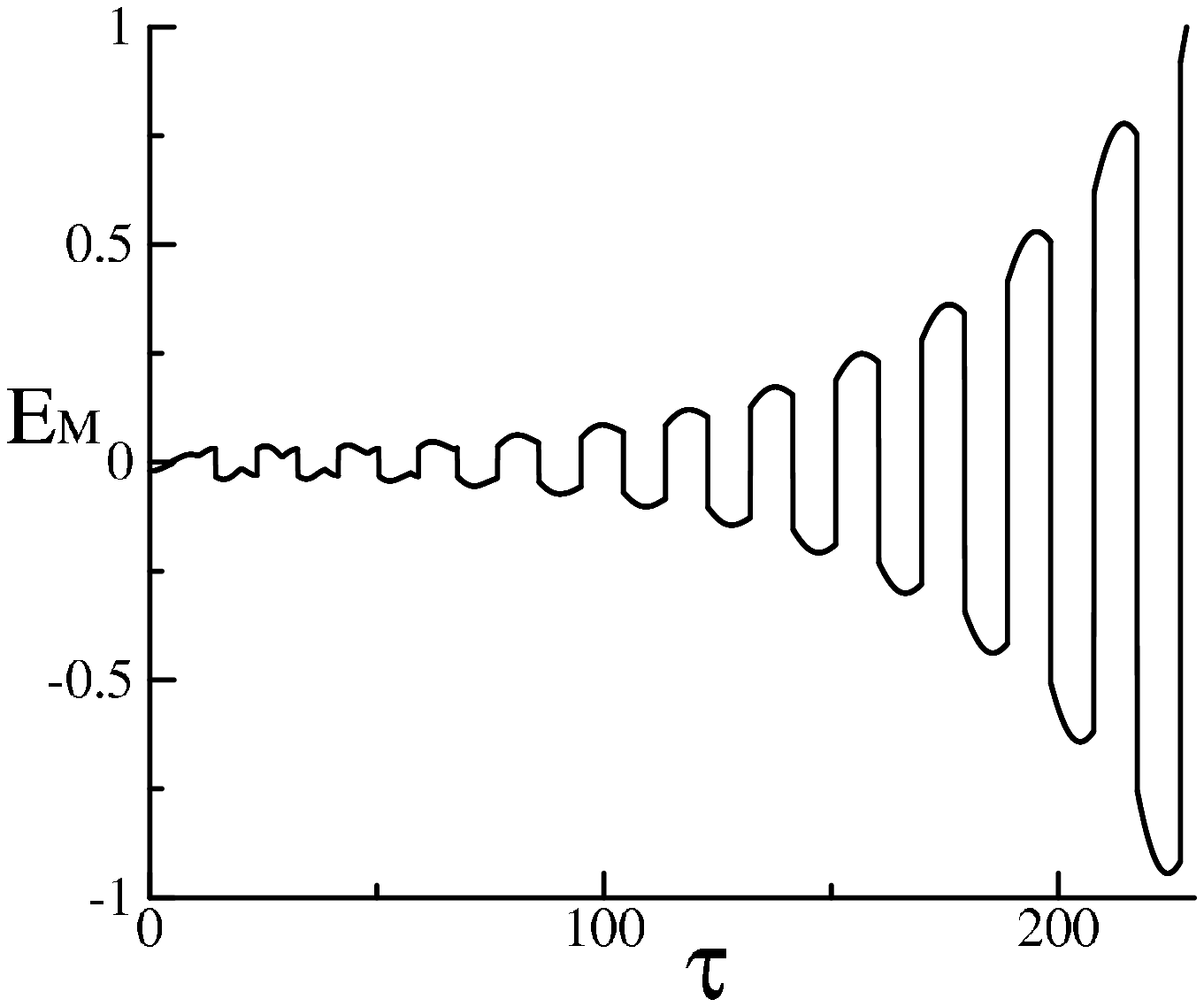}
\caption{Time evolution of $E_{\rm M}$ for RUN1.
}
\end{figure}

\clearpage

\begin{figure}
   \centering
   \includegraphics[width=\textwidth]{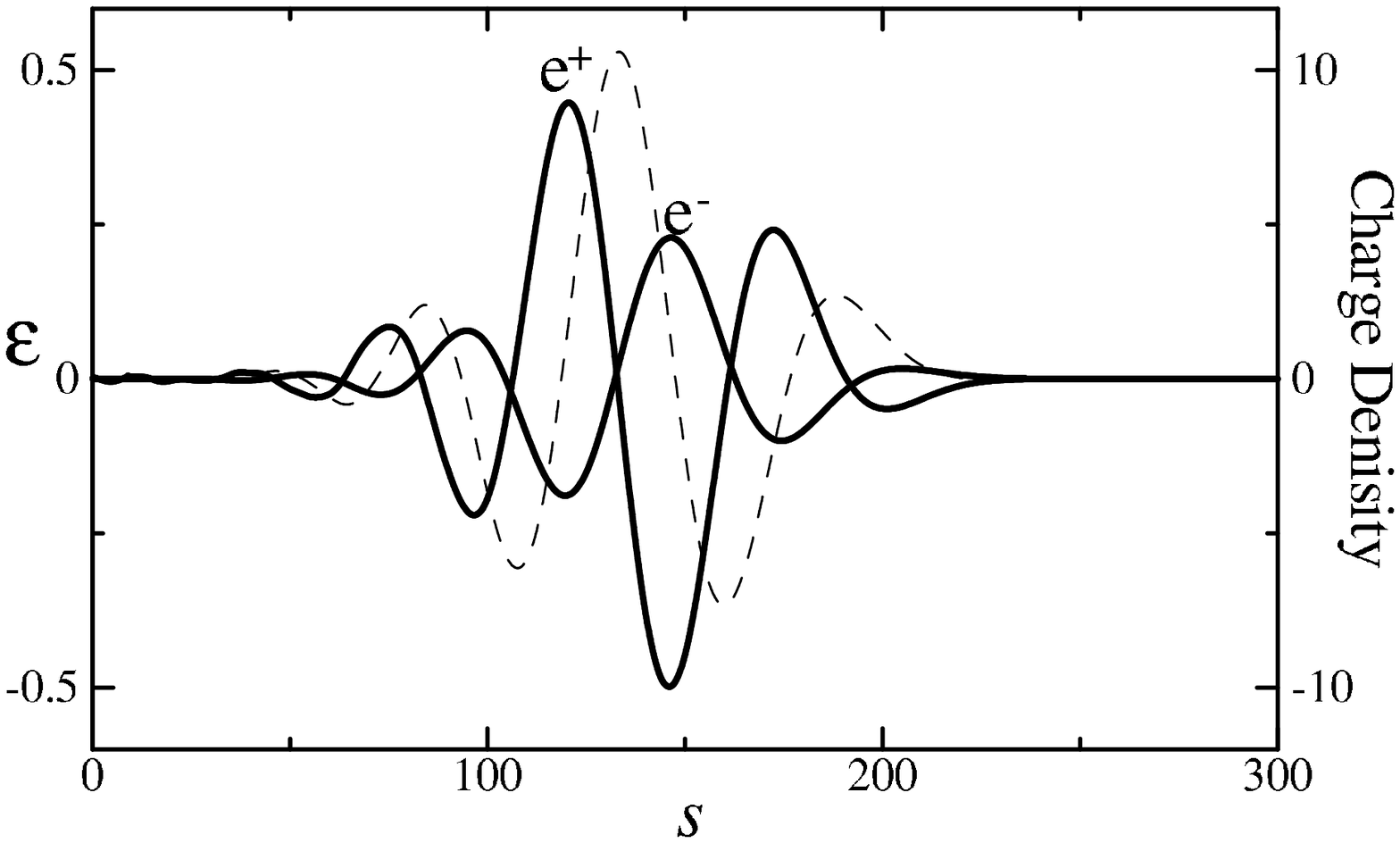}
\caption{Charge-density distributions for $\tau=195$ in RUN1.
The dashed line is the electric field at that time.
}
\end{figure}

\clearpage

\begin{figure}
   \centering
   \includegraphics[width=\textwidth]{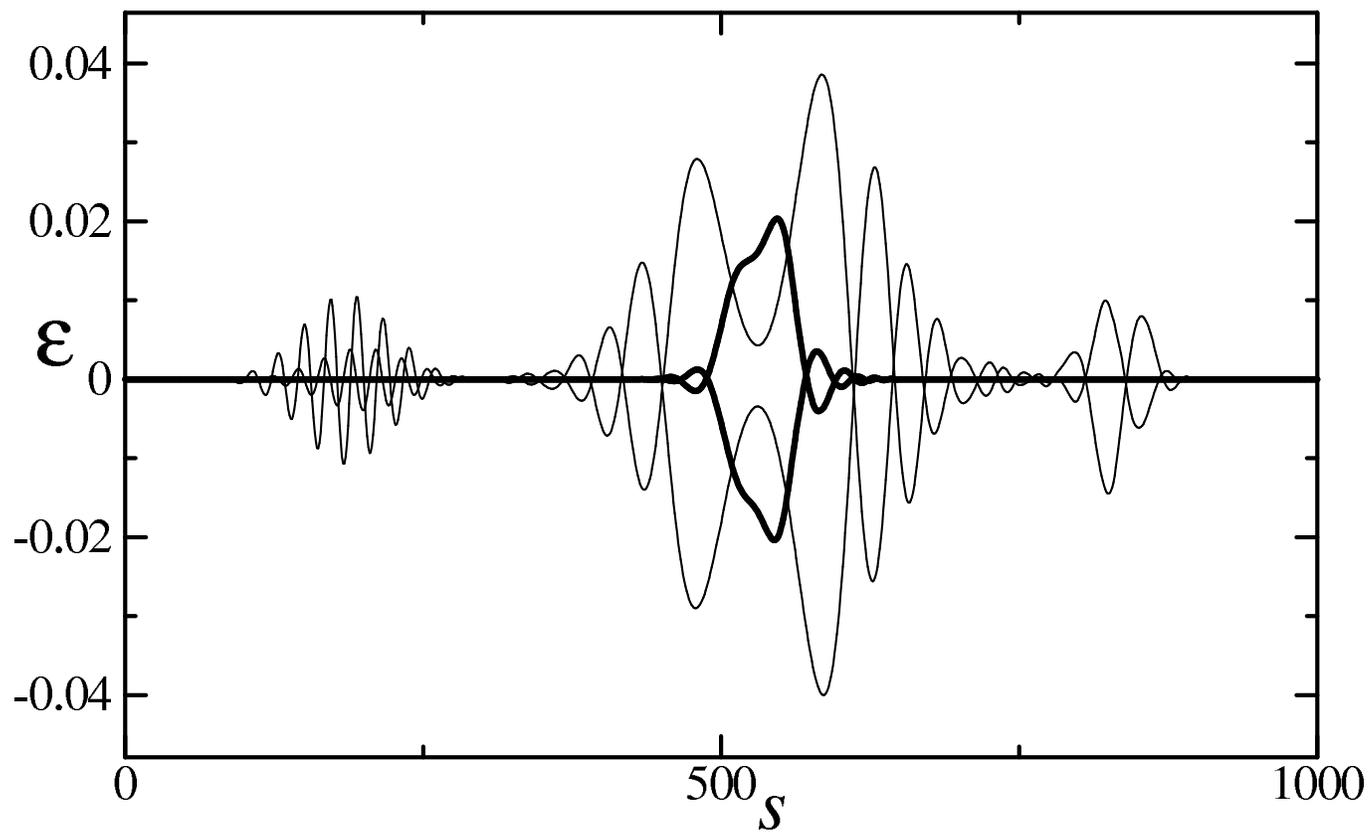}
\caption{Electrostatic waves for RUN2.
The thick lines are for $\tau=80$ and 90.
The thin lines are for $\tau=330$ and 340.
}
\end{figure}

\clearpage

\begin{figure}
   \centering
   \includegraphics[width=\textwidth]{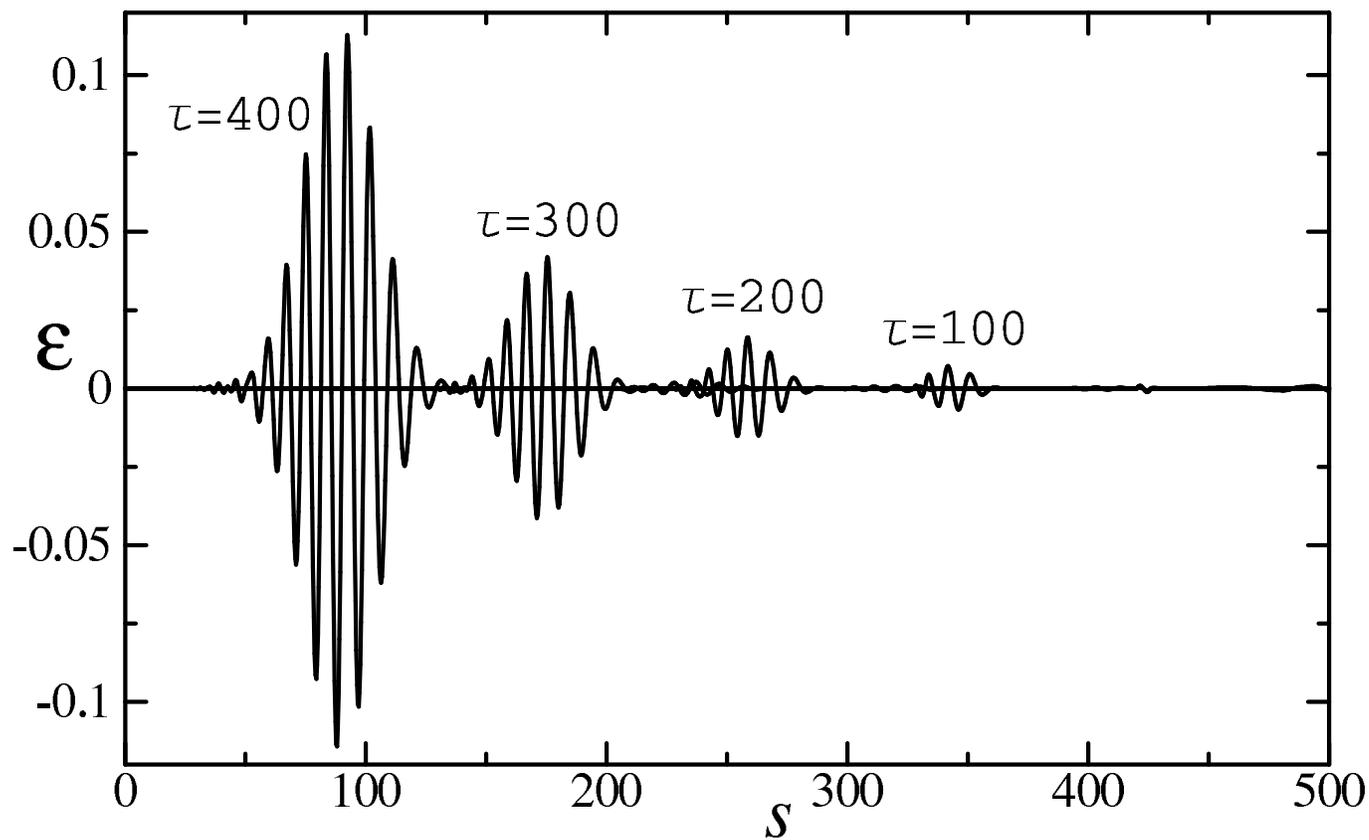}
\caption{Backward waves for RUN3.
}
\end{figure}

\clearpage

\begin{figure}
   \centering
   \includegraphics[width=\textwidth]{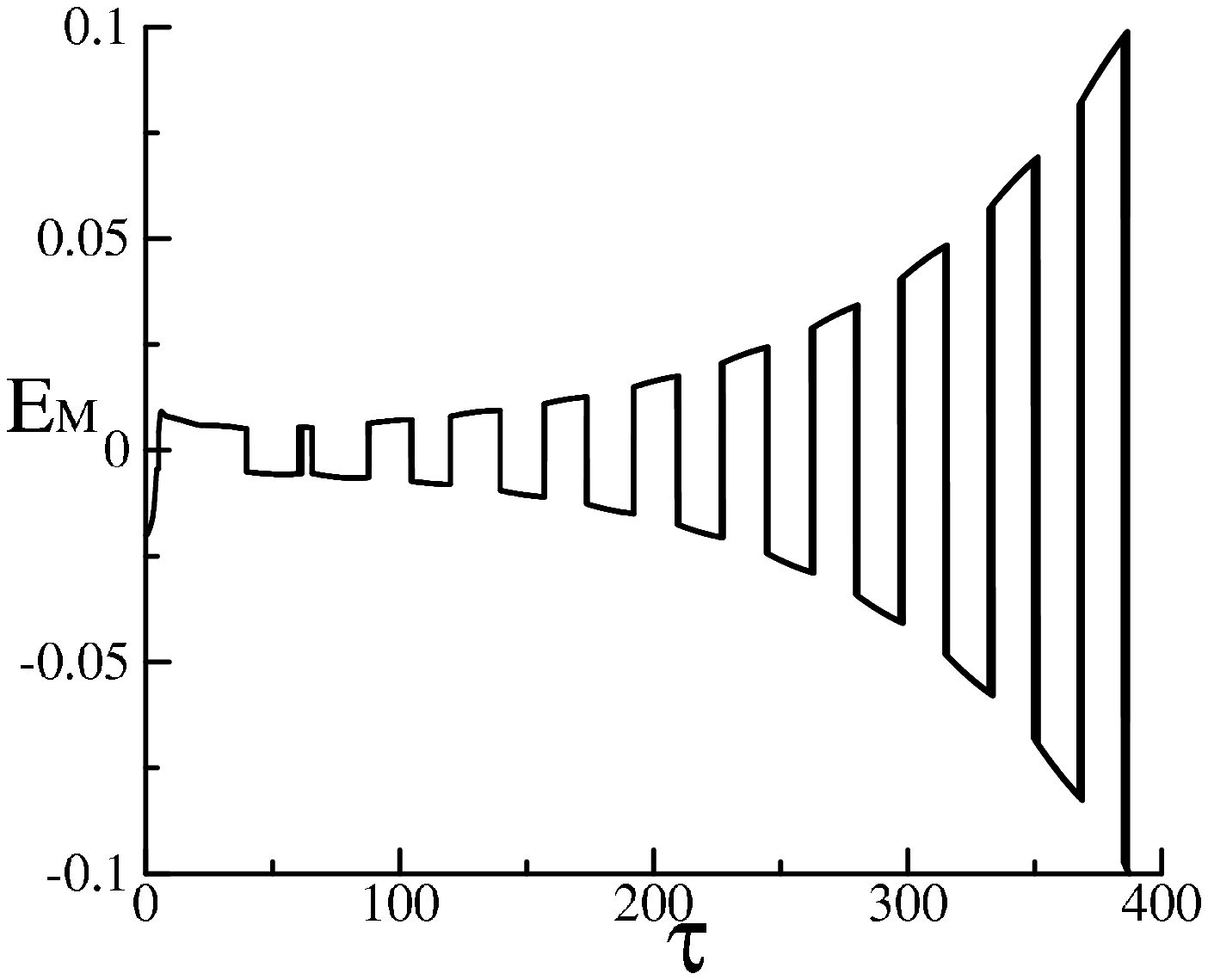}
\caption{Time evolution of $E_{\rm M}$ for RUN3.
}
\end{figure}

\clearpage

\begin{figure}
   \centering
   \includegraphics[width=\textwidth]{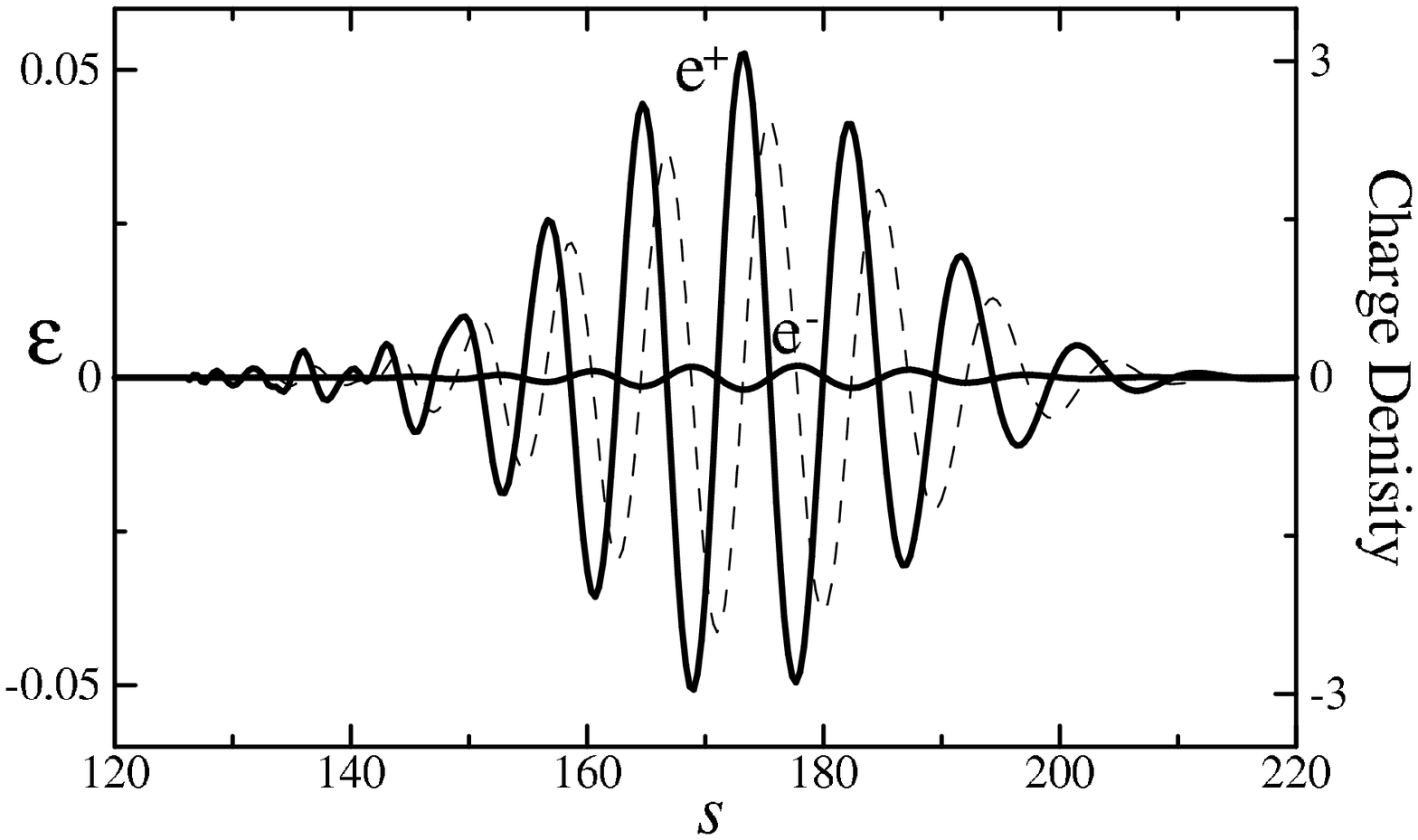}
\caption{Charge-density distributions for $\tau=300$ in RUN3.
The dashed line is the electric field at that time.
}
\end{figure}

\clearpage

\begin{figure}
   \centering
   \includegraphics[width=\textwidth]{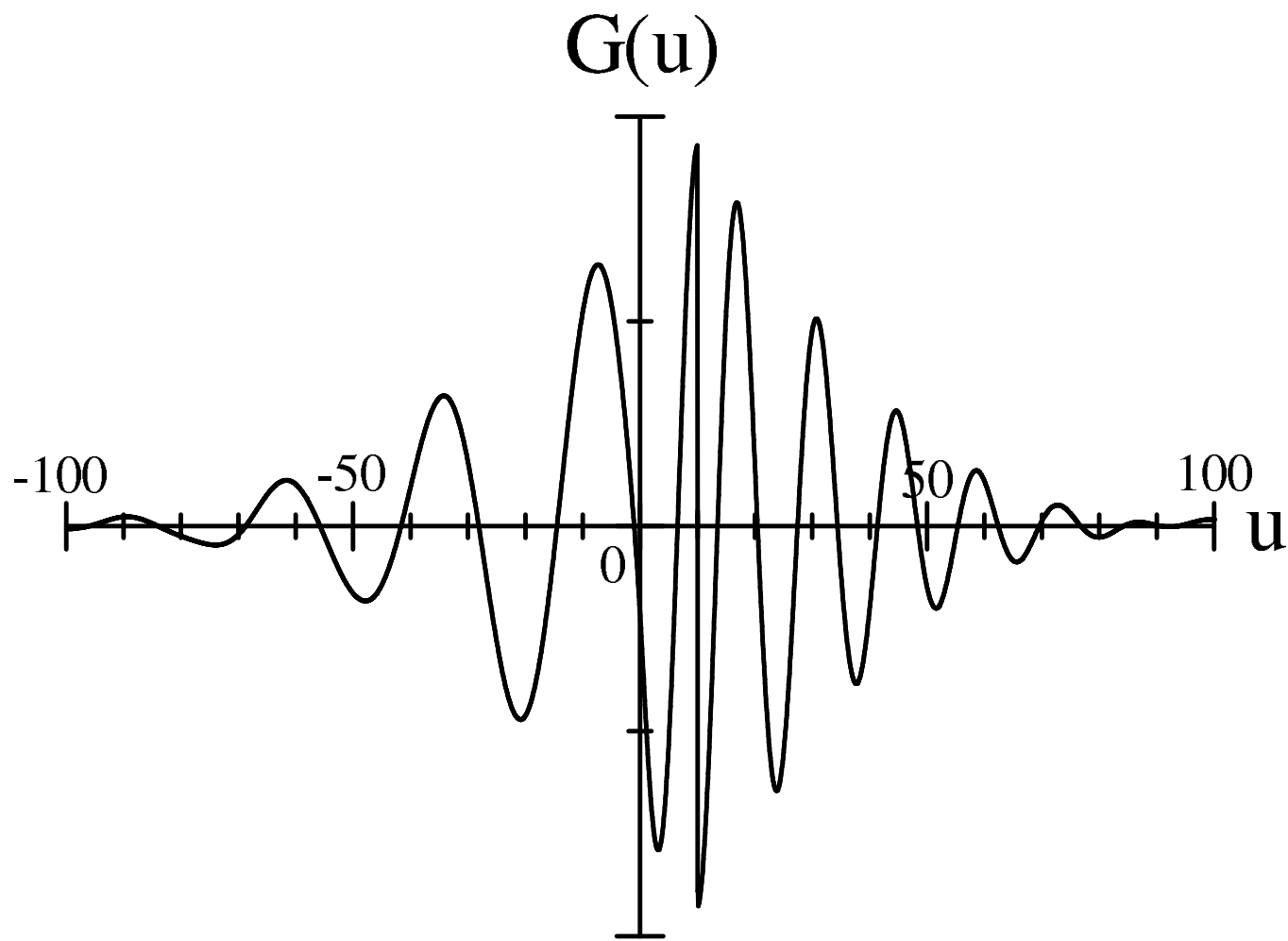}
\caption{The `average force' $G(u)$ for $\tau=195$ in RUN1.
}
\end{figure}

\clearpage

\begin{table}
\begin{tabular}{ccccccc}
\hline \hline
 & $R$ & $u_0$ & $s_M$ & $s_{\rm i}$ & $k_{\rm i}$ & $k_\Delta$ \\ \hline
RUN1 & 0.1 & 10 & 300 & 120 & 0.1 & 1.0 \\
RUN2 & 0.1 & 300 & 1500 & 500 & 0.1 & 0.2 \\
RUN3 & 0.1 & 3 & 500 & 420 & 1.0 & 0.2 \\ 
RUN4 & 0.01 & 100 & 1500 & 200 & 0.01 & 1 \\ 
RUN5 & 0.01 & 30 & 1500 & 1320 & 0.05 & 0.01 \\ 
RUN6 & 0.01 & 3 & 2000 & 1820 & 0.05 & 0.01 \\ \hline
\end{tabular}
\caption{Parameters and initial conditions in computation.}
\end{table}

\begin{table}
\begin{tabular}{cccccc}
\hline \hline
 & $R u_0$ & Type & $t_i \omega_p$ & $\lambda/l_{\rm p}$ & $v_{\rm ph}$ \\ \hline
RUN1 & 1.0 & A & 5 & 6 & -2.8 \\
RUN3 & 0.3 & C & 10 & 0.9 & -1.1 \\ 
RUN4 & 1.0 & A & 5 & 8--10 & -2.8 \\ 
RUN5 & 0.3 & C & 10-30 & 1 & -1.1 \\ \hline
\end{tabular}
\caption{Rough values of the growing time, wavelength, and phase velocity.
The characters ``A'' and ``C'' represent the absolute instability
and convective instability, respectively.}
\end{table}

\end{document}